\documentclass[referee]{raa} 

\usepackage{graphicx,times}
\usepackage{natbib}
\usepackage{amssymb,amsmath}
\bibpunct{(}{)}{;}{a}{}{,}
\usepackage{soul}
\usepackage{xcolor}
\definecolor{dkblue}{RGB}{54, 86, 169}

\newcommand{\FIG}[1] {Figure~\ref{#1} }
\newcommand{\TAB}[1] {Table~\ref{#1} }

\usepackage{fancyhdr}
\usepackage[pagebackref=true]{hyperref}

\begin{document}

 \title{The flat tail in the burst energy distribution of FRB~20240114A}

 \volnopage{ {\bf 20XX} Vol.\ {\bf X} No. {\bf XX}, 000--000}
 \setcounter{page}{1}

 \author{Yu-Xiang Huang\inst{1,2}, Jun-Shuo Zhang\inst{2,3}, Heng Xu\inst{3*}, Long-Fei Hao\inst{1*}, Ke-Jia Lee\inst{4*,1,3,5}, Yong-Kun Zhang\inst{3}, Tian-Cong Wang\inst{6,7}, Shuo Cao\inst{1,2}, De-Jiang Zhou\inst{3}, Jiang-Wei Xu\inst{3}, Zhi-Xuan Li\inst{1}, Yong-Hua Xu\inst{1}, Bo-Jun Wang\inst{3}, Jin-Chen Jiang\inst{3}, Yan-Jun Guo\inst{3}, Zi-Han Xue\inst{4}, Fa-Xin Shen\inst{1,2}, Min Wang\inst{1}, Yun-Peng Men\inst{8}, Wen Chen\inst{1}, Qin Wu\inst{9}, and Fayin Wang \inst{9}
 }

 \institute{ Yunnan Observatories, Chinese Academy of Sciences, Kunming 650216, China; {\it haolongfei@ynao.ac.cn}\\
 \and
 University of Chinese Academy of Sciences, Beijing 100049, China; \\
 \and
 National Astronomical Observatories, Chinese Academy of Sciences, Beijing 100101, China; {\it hengxu@bao.ac.cn}\\
 \and
 Department of Astronomy, Peking University, Beijing 100871, China; {\it kjlee@pku.edu.cn}\\
 \and
 Beijing Laser Acceleration Innovation Center, Beijing 101400, China;\\
 \and 
 School of Physics and Astronomy, Beijing Normal University, Beijing 100875, China;\\
 \and
 Institute for Frontiers in Astronomy and Astrophysics, Beijing Normal University, Beijing 102206, China;\\
 \and
 Max-Planck institut für Radioastronomie, Auf Dem Hügel, Bonn 53121, Germany.\\
  \and
 School of Astronomy and Space Science, Nanjing University, Nanjing 210093, China\\
\vs \no
 {\small Received 20XX Month Day; accepted 20XX Month Day}
}

\abstract{
Fast Radio Bursts (FRBs) are enigmatic millisecond-duration radio transients of extra-galactic origin, whose underlying mechanisms and progenitors remain poorly understood. FRBs are broadly classified into two categories: repeating FRBs, which emit multiple bursts over time, and one-off FRBs, which are detected as single events. A central question in FRB research is whether these two classes share a common origin.
In this study, we present observations of FRB~20240114A, a repeating FRB that entered a hyperactive phase in January 2024. We conducted a 318-hour monitoring campaign using the Kunming 40-Meter Radio Telescope (KM40M) in the S-band (2.187–2.311 GHz), during which we detected eight radio bursts. We analyzed their properties, including dispersion measure (DM), bandwidth, pulse width, flux, fluence, and energy. Additionally, we searched for counterparts in overlapping data from the Five-hundred-meter Aperture Spherical Telescope (FAST) in the L-band (1.0–1.5 GHz). While no bursts were temporally aligned between the two telescopes, we identified one FAST burst that arrived approximately 6 ms after one of the KM40M bursts.
The absence of FAST counterparts for the KM40M bursts suggests that individual bursts from FRB~20240114A are likely narrow-band, with fractional bandwidths less than 10\%. By comparing the cumulative event rates from KM40M and FAST observations, we found that the two measurements are compatible, indicating a possible flattening of the event rate at higher energies. This feature aligns with observations of one-off FRBs, supporting the hypothesis that repeating and one-off FRBs may share a common origin.
}

 \authorrunning{Y.~X. Huang et al.} 
 \titlerunning{The flat tail in the burst energy distribution of FRB~20240114AA} 
 \maketitle
%
\section{Introduction} 
\label{sect:intro}
Fast Radio Bursts (FRBs) are a class of astronomical transient phenomena characterized by extremely bright, millisecond-duration radio pulses of unknown origin \citep{2023RvMP...95c5005Z}. Based on their repeatability, FRBs are classified into two categories: repeating FRBs, which emit multiple bursts over time, and one-off FRBs, which are detected as single events. Since their discovery in 2007 \citep{2007Sci...318..777L}, the number of detected one-off FRBs has grown to approximately 500 \citep{2021ApJS..257...59C}, largely due to large field-of-view radio sky monitoring programs such as CHIME (Canadian Hydrogen Intensity Mapping Experiment\footnote{\url{https://www.chime-frb.ca/}}) and DSA-2000 (Deep Synoptic Array 2000\footnote{\url{https://www.deepsynoptic.org/}}). In contrast, the number of known repeating FRBs remains limited to around 60, despite the detection of tens of thousands of individual bursts from these sources \citep{2022Natur.609..685X,2024ATel16505....1Z}. At present, one of the key questions in FRB research is whether repeating and one-off FRBs share a common origin.


The energy distribution of FRBs has been proposed as a critical observational clue to determine whether the two classes, repeating and one-off FRBs, share a common origin. \citealt{2020MNRAS.494..665L} proposed that one-off FRBs have higher luminosity than that of repeating ones, and \citealt{2024NatAs...8..337K} examined a large population of bursts from the repeating FRB~20201124A and noted that the burst energy distribution exhibits an extended tail with shallower power-law index at higher energies. Notably, this phenomenon has also been observed in other active repeating FRBs, such as FRB~20121102A, FRB~20200120E, FRB~20201124A and FRB~20220912A \citep{2021Natur.598..267L, 2024NatAs...8..337K,2024NatCo..15.7454Z,2024MNRAS_Konijn,2024arXiv241017024O}. These high-energy tails closely resemble the energy distributions observed in non-repeating FRBs, suggesting a possible connection between the two classes. Observing active repeating FRBs can provide detailed insights into the underlying mechanisms of both repeating and one-off FRBs, shedding light on their potential connections and origins.

On January 26, 2024, FRB~20240114A entered an active phase \citep{2024ATel16420....1S}, during which it became hyperactive, emitting bursts at a rate of approximately 500 per hour at a fluence threshold of 15~$\rm mJy\cdot ms$
\citep{2024ATel16505....1Z}. Emission from the source was detected across a wide range of radio frequencies, spanning 0.5–6.0 GHz \citep{2024ATel16494....1P, 2024ATel16620....1L,2024ATel16597....1H,2024ATel16599....1J,2024ATel16547....1P}. This episode of ultra-high activity provided a unique opportunity to investigate whether the energy distribution of repeating FRBs exhibits an extended tail at higher energies, similar to that observed in one-off FRBs.

In this paper, we present observations of FRB20240114A using the KM40M at Yunnan Observatories \citep{Hao2010}. Our findings reveal an extended tail in the burst energy distribution, spanning approximately five orders of magnitude. Detailed descriptions of the observations are provided in \S~\ref{sec:obs}, while the data reduction process and analysis of the energy distribution are discussed in \S~\ref{sec:datred}. Finally, broader implications and conclusions are presented in \S~\ref{sec:con}.

\section{Observation}
\label{sec:obs}
FRB~20240114A was monitored in the S-band (2.187–2.31 GHz) using the KM40M from March 8th, 2024, to November 5th, 2024, with a total observation time of approximately 318 hours. Additionally, the FRB was observed in the L-band (1.0–1.5 GHz) by the FAST \citep{Jiang19SCPMA} from January 28, 2024, to October 29, 2024, accumulating a total observation time of $\sim 36$ hours (see details in Zhang et al., in prep). The overlapping observation period between KM40M and FAST amounted to approximately 7.2 hours. During these observations, the coordinates of FRB~20240114A were adopted from the European VLBI Network (EVN) measurements, with RA =$21^{\mathrm h}27^{\mathrm m}39^{\mathrm s}.835$ and Dec. =$+04^\circ 19'45''.634$ \citep{ATel16542}.

\begin{figure} 
 \centering
 \includegraphics[width=\linewidth, angle=0]{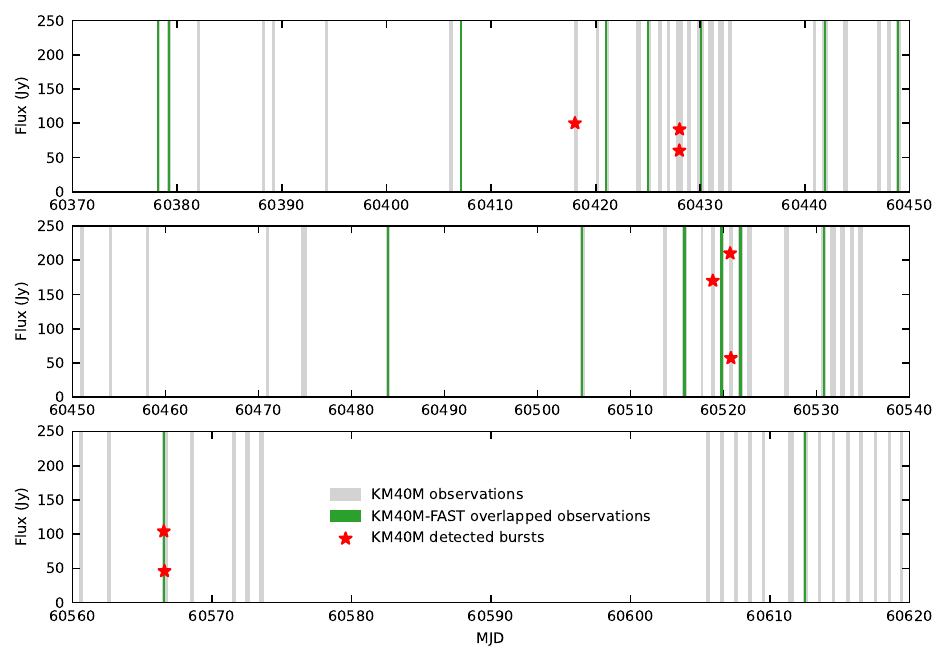}
 \caption{Overview of the Observation Campaign for FRB~20240114A, where
the observation coverage for FRB~20240114A is illustrated. The KM40M observations are represented by gray shaded areas, overlapped observation sessions with FAST are shown in green shaded areas. The flux of bursts detected by KM40M as function of time are plotted with red filled stars with y-axis indicates the burst flux.
 \label{fig:epoch}}
\end{figure}

Prior to each KM40M observation, bright pulsars were observed to verify the stability of the observing system. KM40M observations were conducted in search mode using the Roach2-based digital backend, Burst Emission Automatic Roger (BEAR) \citep{Men2019}. Data were recorded in 8-bit filterbank format with a time resolution of 17.408 $\mu s$. The total observation bandwidth of 124~MHz was divided into 127 channels, providing a frequency resolution of 0.98 MHz. In some of our observation, due to the interference situation, we have to observe with only the left-handed circular polarization. 
The FAST observations were recorded in filterbank format using a \textsc{Roach2}-based digital backend, with a time resolution of 49.152
$\mu s$ and a frequency resolution of 0.122 MHz. A total of 4096 channels were used to cover the 500 MHz bandwidth. The observation coverage is illustrated in \FIG{fig:epoch}.

\section{Data Reduction and Results}
\label{sec:datred}
\subsection{Burst searching and parameter measurements}

We used the FRB search software \textsc{BEAR} \citep{Men2019} to perform an offline search for bursts in the KM40M data. Following the detection of FRB~20240114A by CHIME, which reported a dispersion measure (DM) of 527.7 $\mathrm{pc\;cm^{-3}}$ \citep{2024ATel16420....1S}, we searched for burst signals within a DM range of 522–532 $\mathrm{pc\;cm^{-3}}$ using a step size of 1 $\mathrm{pc\;cm^{-3}}$, which results in maximal 1\% signal-to-noise ratio (S/N) loss for mismatching DM values.

The search employed the matched filter method. The searching grid for pulse width is a geometric series following the equal-signal-to-noise-loss grid approach \citep{Men2019}, where we searched pulse widths ranging from 0.2 ms to 10 ms. We adopted a standard detection threshold of ${\rm S/N}\ge 7.0$, consistent with practices widely used in the FRB community. To mitigate radio frequency interference (RFI), we applied a z-dot filter \citep{Men2019} in the time domain and removed frequency channels with power exceeding the noise floor at the 3-$\sigma$ level.

\begin{figure*}[ht]
\centering
 \includegraphics[width=\linewidth, angle=0]{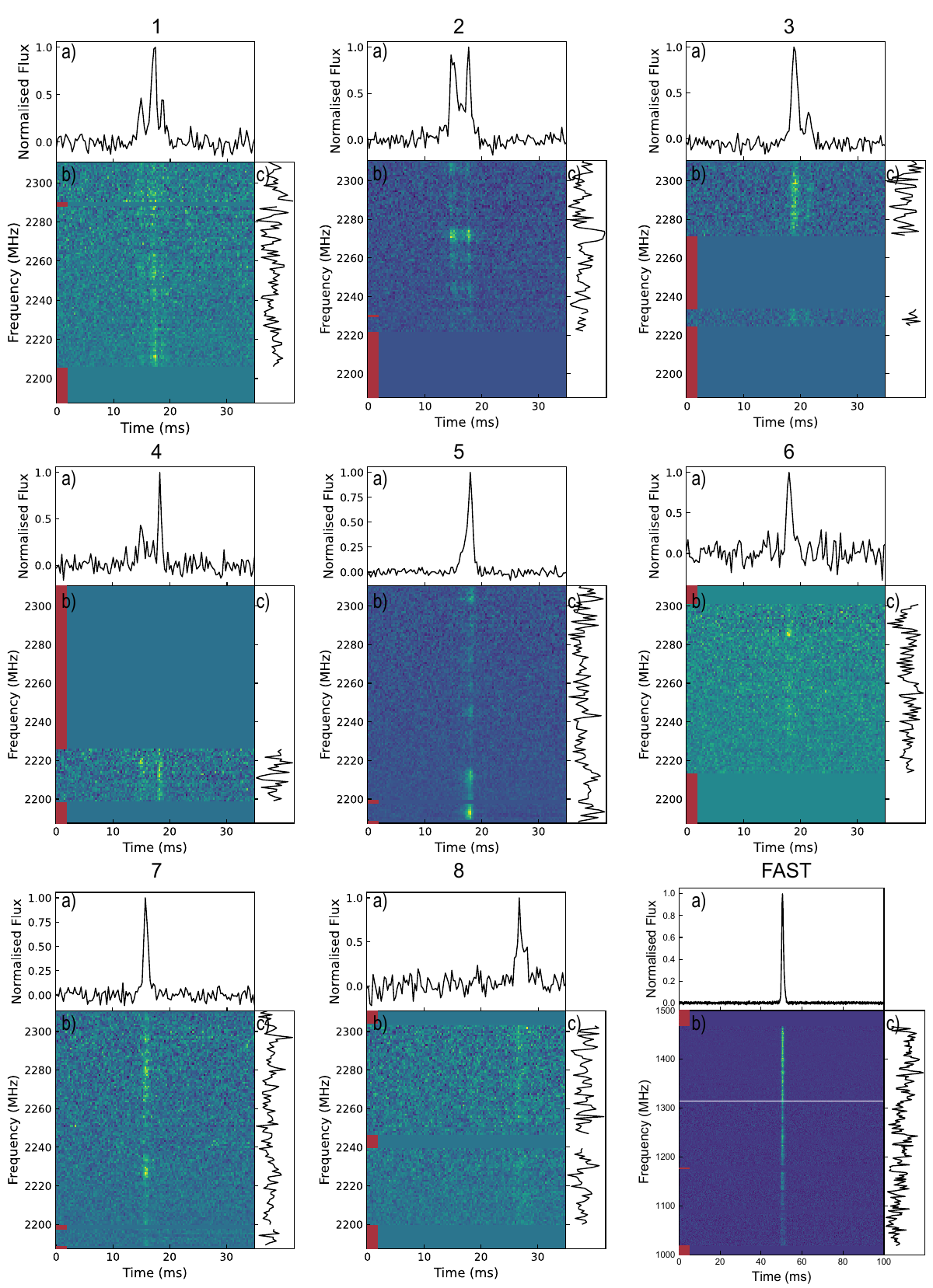}
 \caption{The pulse profiles, dynamic spectra and bandpass of the eight bursts detected with KM40m. We include the dynamic spectrum of one FAST burst here, which arrived 6~ms later than the burst~\#7 of KM40M. The index of each burst is labeled on the top of each panel, where sub-panel a) the pulse profile scaled to peak flux, b) dynamic spectrum with red shade indicating the removed channels due to the RFI, c) bandpass of burst signal. \label{fig:despec}}
\end{figure*}

Each FRB candidate identified by the search pipeline underwent visual inspection to exclude signals affected by the RFI. Candidates exhibiting narrow-band features or zero-DM interference were discarded. After this process, we identified a total of eight radio bursts. Their pulse profiles, bandpass responses, and dynamic spectra are presented in \FIG{fig:despec}. We measured the DM, bandwidth, pulse width, flux, fluence, and energy for each burst. These values are summarized in \TAB{tab:FRB_KM40m}. Since the methods used to derive these parameters are well-established and widely used, we provide only a brief overview of the techniques. Interested readers are encouraged to refer to the cited literature for more detailed information.

\begin{table}[t]
 \centering
 \caption{Parameters of radio bursts detected with KM40M and a potentially associated FAST detection \label{tab:FRB_KM40m}}
 \begin{tabular}{ccccccccc}
 \hline \hline
 \textbf{Burst index} &\textbf{TOA} & \textbf{BW} & \textbf{Width} & \textbf{Flux} & \textbf{Fluence} & \textbf{$E_{\rm iso}$} & \textbf{$\rm DM_{SNR}$}{*} & \textbf{$\rm DM_{opt}$}{\textdagger} \\
 &$\rm MJD$ & $\rm MHz$ & $\rm ms$ & $\rm Jy$ & $\rm Jy\cdot ms$ & $\rm 10^{39}\:erg$ & $\rm pc\cdot cm^{-3}$ & $\rm pc\cdot cm^{-3}$ \\
 \hline
1& 60418.02105159 & 100 &$5.6_{-1.0}^{+1.0}$ &$ 100_{-20}^{+50}$ &$ 190_{-40}^{+100}$ &$ 8.3_{-2}^{+4}$ & $528_{-2}^{+2}$ & $529_{-3}^{+3}$ \\
2\textparagraph& 60427.99986798 & 85 &$5.0_{-0.8}^{+0.8}$ &$ 60_{-12}^{+12}$ &$ 151_{-30}^{+30}$ &$ 5.5_{-1}^{+1}$ & $524_{-3}^{+2}$ & $527_{-13}^{+13}$ \\
3\textparagraph& 60428.01759323 & 46 &$4.7_{-0.9}^{+0.9}$ &$ 91_{-18}^{+18}$ &$ 160_{-30}^{+30}$ &$ 3.2_{-0.6}^{+0.6}$ & $536_{-8}^{+8}$ & $526_{-9}^{+9}$ \\
4& 60518.85123378 & 26  &$5.3_{-1.2}^{+1.2}$ &$ 170_{-30}^{+90}$ &$ 230_{-50}^{+120}$ &$ 2.6_{-0.5}^{+1.4}$ & $539_{-1}^{+1}$ & $560_{-90}^{+90}$ \\
5& 60520.70816271 & 120 &$3.6_{-0.5}^{+0.5}$ &$ 210_{-40}^{+110}$ &$ 270_{-50}^{+140}$ &$ 14_{-3}^{+7}$ & $536_{-1}^{+1}$ & $530_{-2}^{+2}$ \\
6& 60520.79281743 & 85  &$2.8_{-0.7}^{+0.7}$ &$ 57_{-10}^{+30}$ &$ 66_{-13}^{+35}$ &$ 2.4_{-0.5}^{+1.3}$ & $528_{-5}^{+5}$ & $530_{-13}^{+13}$ \\
7& 60566.54379544 & 120 &$2.2_{-0.4}^{+0.4}$ &$ 104_{-21}^{+50}$ &$ 101_{-20}^{+54}$ &$ 5.1_{-1.0}^{+2.7}$ & $528_{-2}^{+2}$ & $528_{-3}^{+3}$ \\
8& 60566.59313771 & 95  &$3.9_{-1.0}^{+1.0}$ &$ 46_{-9}^{+20}$ &$ 86_{-17}^{+46}$ &$ 3.5_{-0.7}^{+1.9}$ & $526_{-2}^{+3}$ & $527_{-8}^{+8}$ \\
FAST\textdaggerdbl\textparagraph & 60566.54380199& 443 &$ 2.3_{-0.1}^{+0.1}$ &$ 1.0_{-0.2}^{+0.2}$ &$ 1.1_{-0.22}^{+0.22}$&$ 0.21_{-0.04}^{+0.04}$ & $529.15_{-0.05}^{+0.19}$ & $528.63_{-0.04}^{+0.04}$\\
 \hline \hline
 \end{tabular}
 \begin{tabular}{l}
 Time of arrival (TOA) is the topocentric value.\\
 * $\rm DM_{\rm SN}$ is the DM derived by maximizing the S/N. \\
 \textdagger $\rm DM_{\rm opt}$ is the DM derived using Fourier domain phase gradient method.\\
 \textdaggerdbl indicates that it is the burst detected in FAST observation, which is only 6~ms later than the burst~\#7. \\
 \textparagraph indicates bursts with dual polarization observation.
 \end{tabular}
\end{table}

The DM value is measured with two standard techniques, where one can either get the DM by maximizing the burst S/N \citep{2012hpa..book.....L} or by aligning the signal phase gradient in Fourier domain \footnote{\url{https://www.github.com/DanieleMichilli/DM_phase}} \citep{2018Natur.553..182M, Xu2024PHD}. Maximum-S/N method is commonly used in pulsar community. However, the method usually produces a biased estimation for the DM due to the frequency evolution of burst profile for FRBs \citep{2019ApJ...876L..23H}. In contrast, Fourier domain phase gradient alignment method was mainly dedicated to FRB signal processing. The method is less affected by the frequency evolution of profile, but tends to yield larger uncertainties. In the current paper, the both of the two methods are used in order to cross-check the results. As shown in \TAB{tab:FRB_KM40m}, all the DM measurements are consistent with each other and with lower-frequency measurements obtained from FAST data.

In this study, the bandwidth ({\rm BW}) of the burst signal is measured \emph{after} the removal of radio frequency interference (RFI). While a common approach involves fitting a Gaussian function to estimate the bandwidth, this method is not well-suited for the KM40M data. The total system bandwidth of KM40M is approximately 100 MHz, and after RFI mitigation, the usable bandwidth is further reduced. Fitting the bandpass with a Gaussian function often results in bandwidth estimates that significantly exceed the frequency range actually covered by the data. Therefore, in this work, we define the bandwidth as the frequency range over which the burst signal remains detectable after RFI mitigation.

To estimate the pulse width, we employed the Gaussian pulse profile fitting method developed in the 1990s \citep{1994A&AS..107..515K}. The FRB pulse profile is fitted with multiple Gaussian functions, where the number of components is incrementally increased until the residuals inside and outside the pulse window exhibit consistent noise properties. Once the parameters of the Gaussian components are determined, the pulse width ($W$) is defined as the time interval during which the flux exceeds 10\% of the peak value. In the pulsar community, this width is commonly referred to as
$W_{\rm 90}$, or the full width at a tenth of the maximum.

We estimate the pulse signal flux using the radiometer equation \citep{2012hpa..book.....L} that
\begin{equation}
 S = \mathrm{S/N} \cdot\frac{T_{\mathrm{sys}}\beta}{G\sqrt{BW \tau n_{\mathrm{p}}}}\,,
\end{equation}
where $S$ is the flux in Jy. $\beta\approx1$ is the digitization factor, and $T_{\mathrm{sys}}\simeq 200~{\rm K}$ is the system temperature of the KM40M room-temperature receiver \citep{Hao2010}. $G=0.27~{\rm K/Jy}$ is the gain of the radio telescope and $BW$ is the burst bandwidth. $\tau$ and $n_{\mathrm{p}}$ are the sampling time and the number of polarization channel, respectively. $\rm S/N$ is the signal to noise ratio of the pulse flux, which is estimated by the ratio between the pulse flux and off-pulse noise root-mean-square (RMS) level. There are two primary sources of error in this flux estimation scheme that 1) fluctuations in $T_{\mathrm{sys}}$
and 2) the right hand circular polarization channel is removed sometimes due to the RFI at KM40M. The fluctuation of $T_{\rm sys}$ leads to 20\% error in flux, and single left-hand circular polarization observation leads to, on average, 50\% underestimation for flux. Thus, the fractional error of flux-related measurements are 20\% and 53\% for lower and upper limits, respectively.

The fluence ($F$) of a radio burst is the time integration of flux ($S$), i.e. $F = \int S\,dt$. Under the assumption that the FRB is isotropic, the energy of the burst \citep{2023RvMP...95c5005Z} is 
\begin{equation}
 E_{\mathrm{iso}} \approx 10^{39} \mathrm{erg} \frac{4\pi}{1+z}\left(\frac{D_{\mathrm{L}}}{10^{28}\;\mathrm{cm}}\right)^{2}\left(\frac{F}{\mathrm{Jy \cdot ms}}\right)\left(\frac{BW}{\mathrm{GHz}}\right)\,,
\end{equation}
where the cosmological redshift and luminosity distance of FRB~20240114A are $z=0.13$ and $D_{\mathrm{L}}=603.7 \,{\rm Mpc}$ \citep{Chen2025ApJL}, respectively. In order to compare the pulse energy with different bandwidth, we compute the spectral energy density, which is defined as $E_{\mathrm{\nu}}=E_{\mathrm{iso}}/{\rm BW}$. 

\begin{figure} 
 \centering
 \includegraphics[width=\linewidth, angle=0]{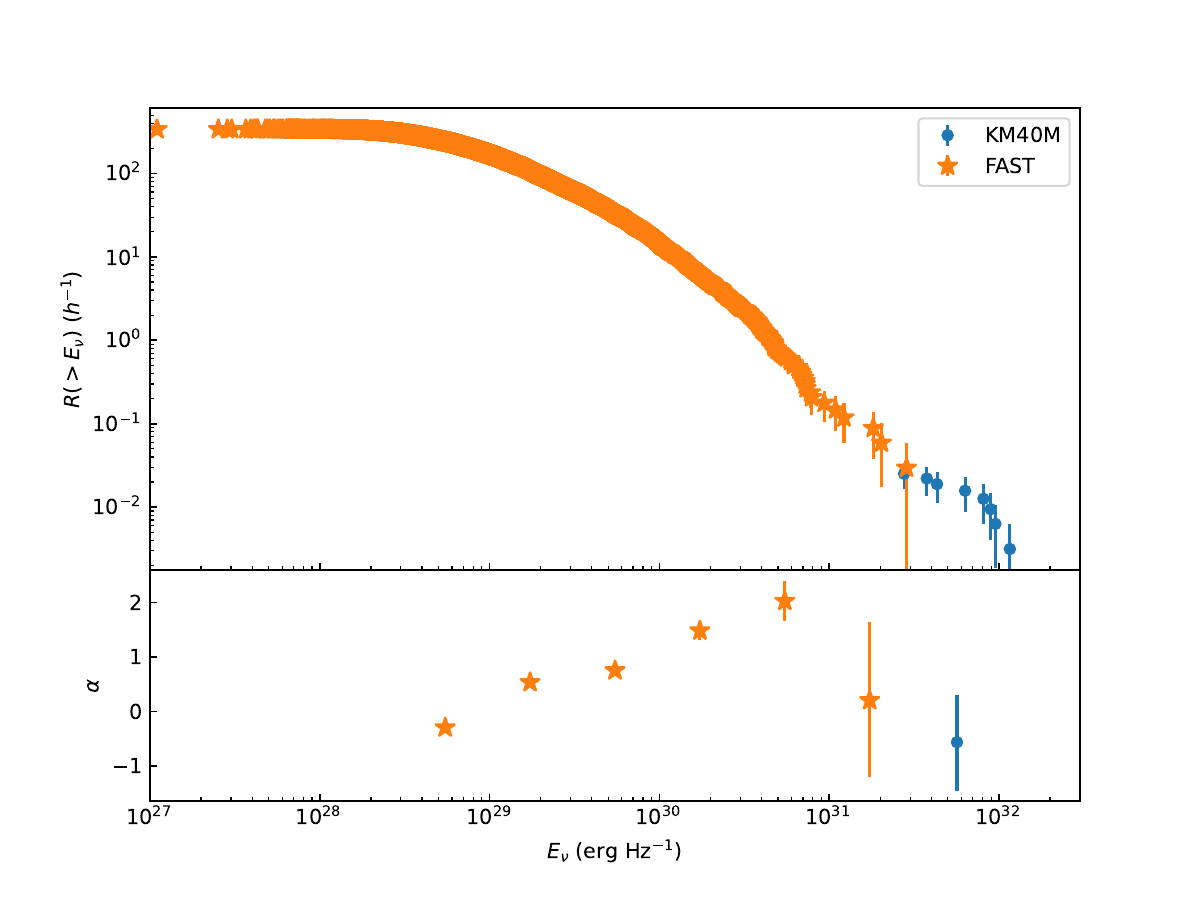}
 \caption{Upper Panel: The cumulative event rate as a function of spectral energy density. The cumulative event rates for bursts detected by FAST and KM40M are included, as indicated in the figure legend. The FAST burst sample will be published in a separate paper (Zhang et al., in preparation).
Lower Panel: The measured power-law index for the cumulative event rate $R(>E_{\nu})$. \label{fig:edf}}
\end{figure}

Using the bursts detected by KM40M, we derived the cumulative event rate as a function of spectral energy density, as shown in \FIG{fig:edf}. To investigate whether the high-energy tail exhibits a different power-law index, we computed the power-law index for the cumulative event rate by fitting windowed data in spectral energy density. Specifically, we modeled the cumulative event rate as $R(>E_{\nu})\propto E_{\nu}^{-\alpha}$, where $\alpha$ is the power-law index. Since the cumulative distribution is intrinsically correlated, directly fitting for the power-law index is statistically inappropriate. Instead, we adopted the multinomial distribution model \citep{1970ApJ...162..405C,2019MNRAS.483.1342J} and used the Bayesian inference to estimate the power-law index and its uncertainty. The software package \textsc{emcee} \citep{2013PASP..125..306F} was employed to perform the posterior sampling. Our results for the power-law index fitting are also presented in \FIG{fig:edf}. The data window for the $\alpha$  spans 0.5 dex in spectral energy density, meaning the upper boundary of the $E_{\nu}$ window is 3 times of the lower boundary (n.b. $\log_{10} 3\simeq 0.5$). 

\subsection{Concurrent burst at 1.0-1.5~GHz observed with FAST}
Thousands of bursts from FRB~20240114A have been detected with FAST. The detailed analysis of the FAST data is beyond the scope of this paper and will be published elsewhere (Zhang et al., in preparation). In this work, we focus on searching for simultaneous bursts in FAST and KM40M data. We searched for FRBs in the FAST observations overlapping with KM40M observations, totaling 7.2 hours of simultaneous coverage. Similar to the KM40M burst search pipeline, we searched radio bursts in the overlapping FAST data within a DM range of 522–532 $\mathrm{pc}\;\mathrm{cm}^{-3}$ and a step size of 0.1 $\mathrm{pc}\;\mathrm{cm}^{-3}$, with a detection threshold of S/N $\ge 7$. Due to the high sensitivity of FAST ($\sim$16~K/Jy \citep{Jiang19SCPMA}), the corresponding flux threshold reaches mJy level (e.g. 4~mJy for 5~ms duration bursts).

\begin{figure} 
 \centering
 \includegraphics[width=7cm, angle=0]{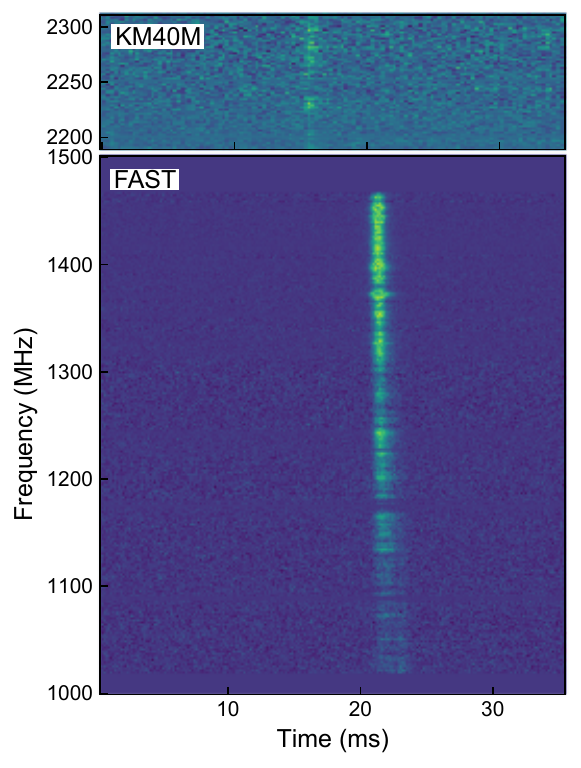} %
 \caption{The dynamic spectra of the burst~\#7 detected with KM40M and the FAST burst 6~ms later. The X-axis is the geocentric burst time, the Y-axis is the observing frequency. \label{fig:kmfast}}
\end{figure}

We relied on the time of arrival (TOA) of bursts to identify the simultaneous detection in FAST and KM40M data. To compare the TOAs of the radio bursts, we used the software package \textsc{TEMPO2} \citep{2006MNRAS.369..655H} to convert topocentric times to geocentric times, correcting for geometric delays and dispersion measure (DM) delays. Previously, simultaneous detections of FRB~20190520B have been achieved using dual-telescope observations with Parkes and FAST\citep{2024ChPhL..41j9501Z}. We tried to search for simultaneous burst with FAST and KM40m and found no bursts perfectly aligned between FAST and KM40M data; however, we identified a radio burst that arrived approximately 6~ms later at FAST than the KM40M burst~\#7. The temporally aligned dynamic spectra of KM40m burst~\#7 and the subsequent FAST burst are shown in \FIG{fig:kmfast}. Both bursts were dedispersed using the DM values that aligned the signal phase gradient in Fourier domain, as listed in \TAB{tab:FRB_KM40m}.

\section{Discussions and Conclusions}
\label{sec:con}

In this paper, we present our efforts to search for radio bursts from FRB~20240114A in the S-band (2.19–2.31 GHz) using the KM40M. Over a total observation time of 318 hours, we detected eight bursts. We also searched for counterparts in overlapping 7.2-hour FAST observations. While no simultaneous counterparts were detected in FAST data, we identified one proximate burst that arrived approximately 6~ms later than the burst~\#7 detected by KM40M. For all bursts, we measured key parameters, including bandwidth, pulse width, flux, fluence, energy, DM, and cumulative event rate, .

It is likely that the \emph{individual} bursts from FRB~20240114A are rather narrow in bandwidth. We can place an upper limit on the intrinsic bandwidth by examining the KM40M detections in the context of FAST non-detections. For example, burst \#7 detected by KM40M reached a flux of 100 to 200~Jy. Assuming a Gaussian-shaped spectrum, the true burst bandwidth must be less than 220~MHz; otherwise, it would have been detected in FAST data. If a power-law spectrum is assumed (i.e. $S\propto f^{\alpha}$),the spectral index must be \emph{positive} and steeper than $\alpha\ge17$, which is unlikely. Thus, we conclude that the individual bursts from FRB~20240114A are narrow-band, with a fractional bandwidth (i.e., the ratio between bandwidth and central frequency) of less than 10\%. The fractional bandwidth of KM40M is only 5\%, so, unfortunately, the current KM40M data is not capable to directly test the bandwidth hypothesis yet.

On the other hand, the spectral density averaged over all bursts from FRB~20240114A may be relatively wide, as indicated by the event rates observed with FAST and KM40M. As shown in \FIG{fig:edf}, the S-band cumulative event rate from KM40M agrees with the L-band results from FAST. If the spectral density of FRB~20240114A evolved significantly from L-band to S-band, we would expect the event rate versus energy density relation to differ markedly. However, there are two caveats to this interpretation. First, the FAST data may be saturated at the higher end of $E_\nu$ (i.e., around $E_\nu \sim 10^{31} \, \rm erg/Hz$), meaning the true event rate of KM40M bursts could extend further into the high $E_\nu$ regime. Second, the KM40M bursts' population is likely incomplete at the lower end of $E_\nu$ (also around $E_\nu \sim 10^{31} \, \rm erg/Hz$), suggesting the true event rate of KM40M bursts could be higher. Considering these caveats, a larger sample of bursts, ideally from long observations with 100-meter-class radio telescopes, is needed to confirm these results. We expect that ultra wideband observations \citep{2024ATel16620....1L} will provide valuable insights.

\begin{figure}
 \centering
 \includegraphics[width=0.66\linewidth, angle=0]{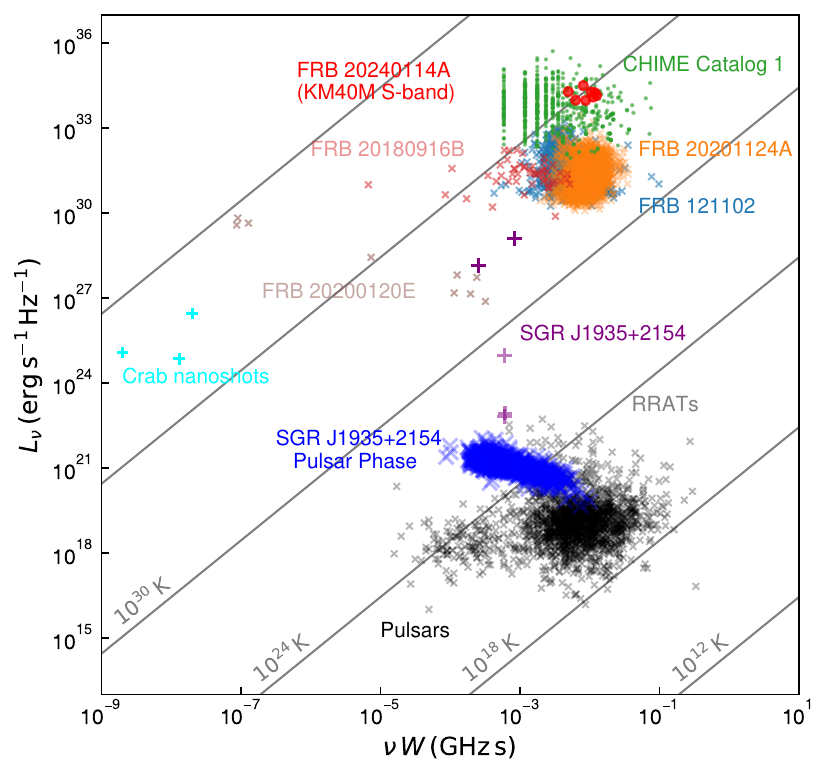}
 \caption{Luminosity density distribution of FRBs and pulsars. The y-axis is the radio luminosity, and the x-axis is the product of observing frequency
and the width of radio emission (bursts or pulses). The emission from pulsars and rotating radio
transients (RRATS) are shown in grey crosses \citep{2005AJ....129.1993M}. The cyan pluses representing Crab nanoshots \citep{2003Natur.422..141H,2007ApJ...670..693H,2010A&A...524A..60J}. Blue cross are for the pulsar phase of the Galactic magnetar SGR~J1935$+$2154 \citep{2023SciA....9F6198Z}. Other symbols are for FRBs from the literature\citep{2020Natur.587...54C,2020Natur.587...59B,2020ATel13699....1Z,2021NatAs...5..414K,2020ATel14080....1P,2022Natur.609..685X,2021Natur.598..267L,2020Natur.582..351C,2021NatAs...5..594N,2022NatAs...6..393N} and the CHIME/FRB Catalog 1 \citep{2021ApJS..257...59C}.}
 \label{fig:vw-lv}
\end{figure}

We measured the power-law index for the event rate versus energy density relation, as shown in \FIG{fig:edf}. We noted features similar to those observed in previous studies: the power-law index increases as a function of burst energy for $E_\nu \in [3 \times 10^{28}, 10^{31}]$~erg/Hz. However, the power-law index appears to drop suddenly to smaller values at
$E_\nu \sim 10^{31}$~erg/Hz, although the statistical significance of this feature is modest (
$\sim 3$-$\sigma$ confidence) in our data. Previous investigations of active repeating FRBs \citep{2023MNRAS.519..666J,2024NatAs...8..337K,2024NatCo..15.7454Z,2024MNRAS_Konijn,2024arXiv241017024O} have reported similar features, suggesting that the flat tail in the event rate at higher burst energies may be a universal characteristic of repeating FRBs. These rare high-energy bursts may form a population analogous to one-off FRBs, as previously argued. Indeed, when comparing the luminosity density of FRB~20240114A measured with KM40M to other known FRBs and pulsars, the KM40M detections overlap well with one-off FRBs, the majority of sources in the CHIME/FRB Catalog 1, as shown in \FIG{fig:vw-lv}. Compared to pulsars, another class of celestial objects that emit repeating radio pulses, bursts detected by KM40M further widen the observed gap between FRBs and radio pulsars.
The prevailing theories attributes FRB emission to magnetar magnetospheres, supported by substantial observational evidence \citep{2020Natur.586..693L,2020Natur.587...59B,2020Natur.587...54C,2022Natur.609..685X,2024NSRev..12..293J}. The extreme flux difference between FRBs and pulsar emissions (spanning 13 orders of magnitude as shown in \FIG{fig:vw-lv}) significantly intensifies the long-standing question of how pulsar-like magnetospheres can repeatably produce such extraordinarily luminous radio bursts.

For the proximate burst of KM40M burst~\#7 detected in FAST data, it is unlikely that the 6~ms delay is due to a misalignment in burst timing. Both the FAST and KM40M positions are accurate to within 10 meters, making it improbable that the 6~ms difference is caused by errors in geometric delay, which would require an implausible 1800~km error in telescope position. Additionally, the BEAR system uses the local network time service, with an error of at most 2~ms, which cannot account for the observed delay. Finally, errors in the dispersion measure (DM) do not contribute significantly to the 6~ms delay. In the work, we have corrected the time delay between 1.5~GHz (FAST) and 2.187~GHz (KM40M) by $\Delta T=1.0 {\rm ms\,} {\rm DM/(cm^{-3}\,pc)}\simeq 0.5\,{\rm s}$.
The DM error for the FAST burst is approximately 0.2~$\rm pc\cdot cm^{-3}$, which introduces, at most, a 0.2~ms error in the time difference between L-band (FAST) and S-band (KM40M). Thus, the significance level of 6~ms delay between FAST and KM40m burst is 25-$\sigma$ given the DM error.
On the other hand, it is possible that the two bursts separated by 6~ms belong to a single burst, with the 6~ms delay representing the time-frequency downward drifting timescale \citep{2019ApJ...876L..23H}. The ultra wideband observations \citep{2024ATel16620....1L} will provide further insights into such wideband time-frequency downward drifting phenomena, helping to differentiate between models of time-frequency downward drifting mechanisms \citep{2019ApJ...876L..15W,2021arXiv210713549T}.

\normalem
\begin{acknowledgements}
This work is supported by the National SKA Program of China (Grant No. 2020SKA0120100), the Special Project of Foreign Science and Technology Cooperation, Yunnan Provincial Science and Technology Department (Grant No. 202003AD150010), the National Key R\&D Program of China (Grant No. 2022YFC2205203), the National Natural Science Foundation of China (NSFC, Grant No. 12073076, 12173087, 12041303,
and 12063003), the CAS ’Western Light Youth Project’,  Yunnan Fundamental Research Projects(Grant No. 202401AT070144 and 202505AO120021) and funding from the Max-Planck Partner Group. K.J.~Lee acknowledges support from the XPLORER PRIZE.
\end{acknowledgements}
 
\bibliographystyle{raa}
\bibliography{ref}

\end{document}